\documentclass[a4paper,twoside,12pt]{article}
\input{obsjkt.sty}

\newcommand{\Msun}{\ensuremath{~{\rm M}_\odot}}                   
\newcommand{\Rsun}{\ensuremath{~{\rm R}_\odot}}                   
\newcommand{\rhosun}{\ensuremath{~\rho_\odot}}                    
\newcommand{\Teff}{\ensuremath{T_{\rm eff}}}                      
\newcommand{\degr}{\ensuremath{^\circ}}                           
\renewcommand{\kms}{~km~s$^{-1}$}                                 
\newcommand{\as}{\ensuremath{^{\prime\prime}}}                    

\newcommand{\etal}{\textit{et al.}}                               

\newcommand{\tess}{\textit{TESS}}
\newcommand{\hip}{\textit{Hipparcos}}
\newcommand{\gaia}{\textit{Gaia}}

\newcommand{\debcat}{\textit{DEBCat}}
\newcommand{\targ}{V1388~Ori}

\newcommand{\Msunnom}{\hbox{$\mathcal{M}^{\rm N}_\odot$}}
\newcommand{\Rsunnom}{\hbox{$\mathcal{R}^{\rm N}_\odot$}}
\newcommand{\Lsunnom}{\hbox{$\mathcal{L}^{\rm N}_\odot$}}

\usepackage{rotating}

\begin{document} 

\OBSheader{Rediscussion of eclipsing binaries: \targ}{J.\ Southworth \& D.\ M.\ Bowman}{2022 September}

\OBStitle{Rediscussion of eclipsing binaries. Paper X. \\ The pulsating B-type system V1388 Orionis}

\OBSauth{John Southworth$^1$ and Dominic M.\ Bowman$^2$}

\OBSinst{Astrophysics Group, Keele University, Staffordshire, ST5 5BG, UK}
\OBSinst{Institute of Astronomy, KU Leuven, Celestijnenlaan 200D, B-3001 Leuven, Belgium}

\OBSabstract{\targ\ is an early-B type detached eclipsing binary whose physical properties have previously been measured from dedicated spectroscopy and a ground-based survey light curve. We reconsider the properties of the system using newly-available light curves from the Transiting Exoplanet Survey Satellite (\tess). We discover two frequencies in the system, at 2.99~d$^{-1}$ and 4.00~d$^{-1}$ which are probably due to $\beta$~Cephei or slowly-pulsating B-star pulsations. A large number of additional significant frequencies exist at multiples of the orbital frequency, 0.4572~d$^{-1}$. We are not able to find a fully satisfactory model of the eclipses, but the best attempts show highly consistent values for the fitted parameters. We find masses of $7.24 \pm 0.08$\Msun\ and $5.03 \pm 0.04$\Msun, and radii of $5.30 \pm 0.07$\Rsun\ and $3.14 \pm 0.06$\Rsun. The properties of the system are in good agreement with the predictions of theoretical stellar evolutionary models and the \gaia\ EDR3 parallax if the published temperature estimates are revised downwards by 1500~K, to 19\,000~K for the larger and more massive star and 17\,000~K for its companion.}


\section*{Introduction}

Detached eclipsing binaries (dEBs) are our primary source of measurements of the physical properties of normal stars \cite{Andersen91aarv,Torres++10aarv,Me15aspc}. Those containing high-mass stars are of particular importance because such stars dominate the light of young stellar populations \cite{Demello++00apj,Robertson+10nat} and the chemical evolution of galaxies \cite{Langer12araa}, and give rise to a wide variety of exotic objects \citep{Podsiadlowski+02apj,Podsiadlowski+04apj,Belczynski+20aa,Chrimes+20mn}. Theoretical models of massive stars remain limited by the imperfect understanding of several phenomena including internal mixing and convective core overshooting \citep{Tkachenko+20aa,Johnston21aa}, angular momentum transport \cite{Aerts++19araa} and the effects of internal gravity waves \cite{Bowman+19aa,Bowman+20aa}. Massive stars are typically found in multiple systems \cite{Sana+14apjs,Kobulnicky+14apjs}, and their evolution is dominated by binary interactions \cite{Sana+12sci}.

In this work we revisit the \targ\ system, using a recently-obtained space-based light curve, with the aim of determining its physical properties to high precision \cite{Me20obs}.  Its spectrum was classified as B2~V by Walborn \cite{Walborn71apjs} and it has been used as a spectral standard star \cite{WalbornFitzpatrick90pasp}, before the discovery of eclipses in its light curve from the \hip\ satellite \cite{Hipparcos97,Kazarovets+99ibvs}. A detailed study of \targ\ was presented by Williams \cite{Williams09aj} (hereafter W09) based on 29 coud\'e spectra (resolving power $R=11\,500$) and a scattered $V$-band light curve from the All Sky Automated Survey (ASAS, Pojma\'nski \cite{Pojmanski97aca,Pojmanski02aca}). W09 used the {\sc elc} code \cite{OroszHauschildt00aa} to fit the light and radial velocity (RV) curves and measure the properties of the system. He also determined effective temperature (\Teff) values of $20\,500 \pm 500$~K and $18\,500 \pm 500$~K for the two stars from the tomographically-reconstructed spectra of the individual stars.


Table~\ref{tab:info} contains basic information for \targ. The $BV$ magnitudes are from the Tycho mission \cite{Hog+00aa} and are evenly distributed in orbital phase, so represent the average magnitude of the system. The 2MASS $JHK_s$ magnitudes were obtained at a single epoch corresponding to an orbital phase of $0.4081 \pm 0.0012$, so represent the brightness of the system shortly before the start of secondary eclipse.

\begin{table}[t]
\caption{\em Basic information on \targ. \label{tab:info}}
\centering
\begin{tabular}{lll}
{\em Property}                            & {\em Value}                 & {\em Reference}                   \\[3pt]
Henry Draper designation                  & HD 42401                    & \cite{CannonPickering18anhar2}    \\
\textit{Hipparcos} designation            & HIP 29321                   & \cite{Hipparcos97}                \\
\textit{Tycho} designation                & TYC 738-244-1               & \cite{Hog+00aa}                   \\
\tess\ Input Catalog designation          & TIC 337165095               & \cite{Stassun+19aj}               \\
\textit{Gaia} EDR3 designation            & 3342421035256268544         & \cite{Gaia21aa}                   \\
\textit{Gaia} EDR3 parallax               & $1.3198 \pm 0.0432$ mas     & \cite{Gaia21aa}                   \\          
$B$ magnitude                             & $7.424 \pm 0.015$           & \cite{Hog+00aa}                   \\          
$V$ magnitude                             & $7.493 \pm 0.018$           & \cite{Hog+00aa}                   \\          
$J$ magnitude                             & $7.506 \pm 0.024$           & \cite{Cutri+03book}               \\
$H$ magnitude                             & $7.541 \pm 0.027$           & \cite{Cutri+03book}               \\
$K_s$ magnitude                           & $7.551 \pm 0.034$           & \cite{Cutri+03book}               \\
Spectral type                             & B2 V                        & \cite{Walborn71apjs}              \\[3pt]
\end{tabular}
\end{table}


\section*{Observational material}

\begin{figure}[t] \centering \includegraphics[width=\textwidth]{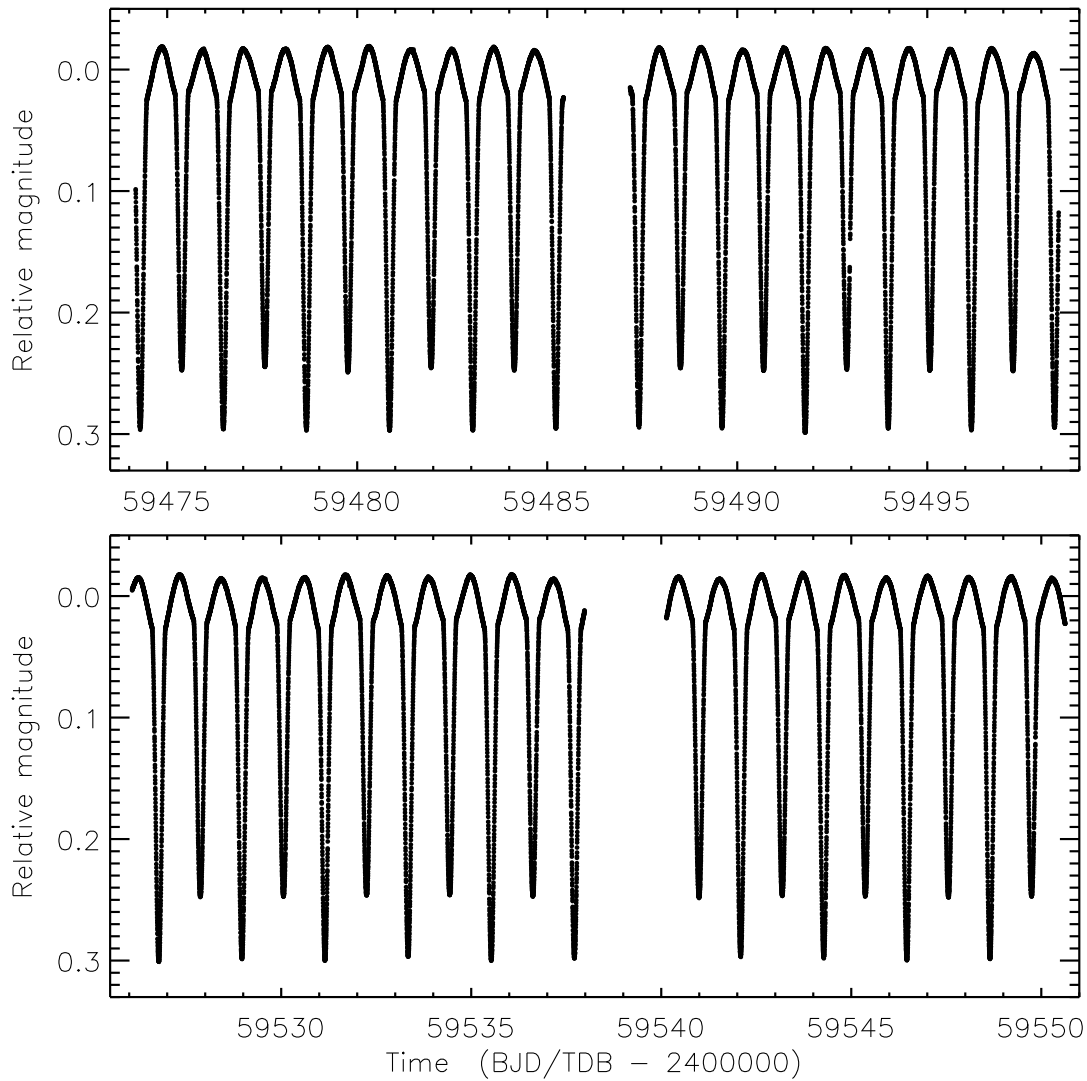} \\
\caption{\label{fig:time} \tess\ short-cadence SAP photometry of \targ\ from sectors 43 (top
panel) and 45 (bottom panel). The flux measurements have been converted to relative magnitude
and rectified to zero magnitude by subtraction of low-order polynomials.} \end{figure}


\targ\ has been observed on three occasions by the NASA \tess\ satellite \cite{Ricker+15jatis}. It was observed in long cadence (600~s sampling rate) in sector 33 (2020/12/17 to 2021/01/13) but we did not use these data due to their coarser temporal sampling. It was observed in short cadence (120~s sampling rate) in sectors 43 (2021/09/16 to 2021/10/12) and 45 (2021/11/06 to 2021/12/02), and these data were analysed for this work.

We downloaded the data from the MAST archive\footnote{Mikulski Archive for Space Telescopes, \\ \texttt{https://mast.stsci.edu/portal/Mashup/Clients/Mast/Portal.html}} and converted the fluxes to relative magnitude. We retained only those observations with a QUALITY flag of zero, leaving 32\,307 of the original 35\,893 datapoints. The simple aperture photometry (SAP) and pre-search data conditioning SAP (PDCSAP) data \cite{Jenkins+16spie} were visually almost indistinguishable, the only clear differentiating feature being a 0.002~mag variation in eclipse depth. We therefore adopted the SAP data as usual in this series of papers, and ensured that we fitted for third light in the analysis below. The light curve is shown in Fig.~\ref{fig:time}.


\subsection*{Initial analysis of the light curve}

The components of \targ\ are significantly distorted due to their large fractional radii ($r_{\rm A}$ and $r_{\rm B}$ where $r_{\rm A} = {R_{\rm A}}/{a}$, $r_{\rm B} = {R_{\rm B}}/{a}$, $R_{\rm A}$ and $R_{\rm B}$ are the radii of the stars, and $a$ is the semimajor axis of the relative orbit) so the light curve must be analysed using a model incorporating Roche geometry. However, this is time-intensive due to the large calculation time required to model a large number of datapoints using existing Roche-geometry codes. We therefore performed a preliminary analysis with the {\sc jktebop}\footnote{\texttt{http://www.astro.keele.ac.uk/jkt/codes/jktebop.html}} code \cite{Me++04mn2,Me13aa} in order to determine the orbital ephemeris of the system, thus allowing us to convert all the data to orbital phase and bin into a small number of phased datapoints.

We fitted the full \tess\ sector 43 and 45 light curve with {\sc jktebop}, assuming a circular orbit. We defined star~A to be that eclipsed at primary minimum, making it the hotter of the two components (and also the larger and more massive component in the case of \targ), and star~B to be its companion. The orbital ephemeris resulting from this fit is
\begin{equation}
  \mbox{Min~I} = \mbox{BJD/TDB } 2459485.226609 (16) + 2.18703006 (82) E              
\end{equation}
where $E$ is the cycle number since the reference time and the bracketed quantities indicate the uncertainties in the last digit of the preceding number.

We then sought to check this against a timing from W09. Only one timing is given by W09\footnote{W09 quote their reference time using the notation $T_{\rm IC,1}$ to denote the ``time of inferior conjunction of the primary star''. However, they also note that ``this is the time of secondary minimum in the light curve''. The two statements are in mutual conflict. Using our ephemeris we determined that the $T_{\rm IC,1}$ value in W09 is indeed a time of secondary eclipse, and therefore it is a time of \emph{superior} conjunction of star~A.} (their table~3), which we converted from the UTC timescale to TDB using the IDL routines of Eastman \etal\ \cite{Eastman++10pasp}. The timing corresponds to an orbital phase of $0.5492 \pm 0.0009$, which is sufficiently distant from a time of eclipse to suggest that the orbital period of \targ\ may not be constant.

Using the ephemeris above, we converted each of the \tess\ sector 43 and 45 light curves into orbital phase and binned them into 400 points equally distributed over phases 0 to 1. The two binned light curves look almost identical. As a check, we phase-binned the PDCSAP data as well, and found that the slightly larger eclipse depths in the PDCSAP data occurred in both \tess\ sectors.


\subsection*{Frequency analysis}

Following previous work on massive pulsators in eclipsing binaries \cite{MeBowman22mn}, the residuals of the {\sc jktebop} fit to the unbinned SAP data were used to search for the presence of pulsations in \targ. We used the combined sectors 43 and 45 residual light curve and calculated the discrete Fourier transform \cite{Kurtz85mn}. There are many significant peaks in the resultant amplitude spectrum, but almost all of them fall at integer multiples of the orbital frequency. Therefore, they probably do not represent independent pulsation mode frequencies, but are rather a consequence of an imperfect binary model leaving residual signal at orbital harmonics (see below).

However, we detected two significant frequencies that do not coincide with orbital harmonics, namely $2.9943 \pm 0.0002$~d$^{-1}$ and $3.9987 \pm 0.0004$~d$^{-1}$. We define \emph{significant} to mean that the signal-to-noise ratio (S/N) is larger than five in the amplitude spectrum after all orbital harmonics have been removed, using a 1~d$^{-1}$ frequency window in the amplitude spectrum centred on the extracted frequency to estimate the local noise level. The amplitudes of the two signals are 0.33 and 0.18 mmag, respectively. Additional variability at lower frequency is visually evident, in particular at 1.0 and 2.0~d$^{-1}$, but has S/N $<$ 4 so formally falls below our detection threshold and is not significant.

These frequencies are typical of $\beta$~Cephei or SPB pulsation modes \cite{StankovHandler05apjs,Waelkens91aa,Aerts++10book}. The spectral type of \targ\ is consistent with such pulsations, so we conclude that the system is a new example of a massive pulsator in a dEB. Such systems are rare and could provide useful probes of stellar interiors through forward asteroseismic modelling. The closeness of these two frequencies to multiples of the Earth's rotational frequency is unlikely to be meaningful because the TESS data are not ground-based.

\begin{sidewaysfigure} \centering
\includegraphics[width=\textwidth]{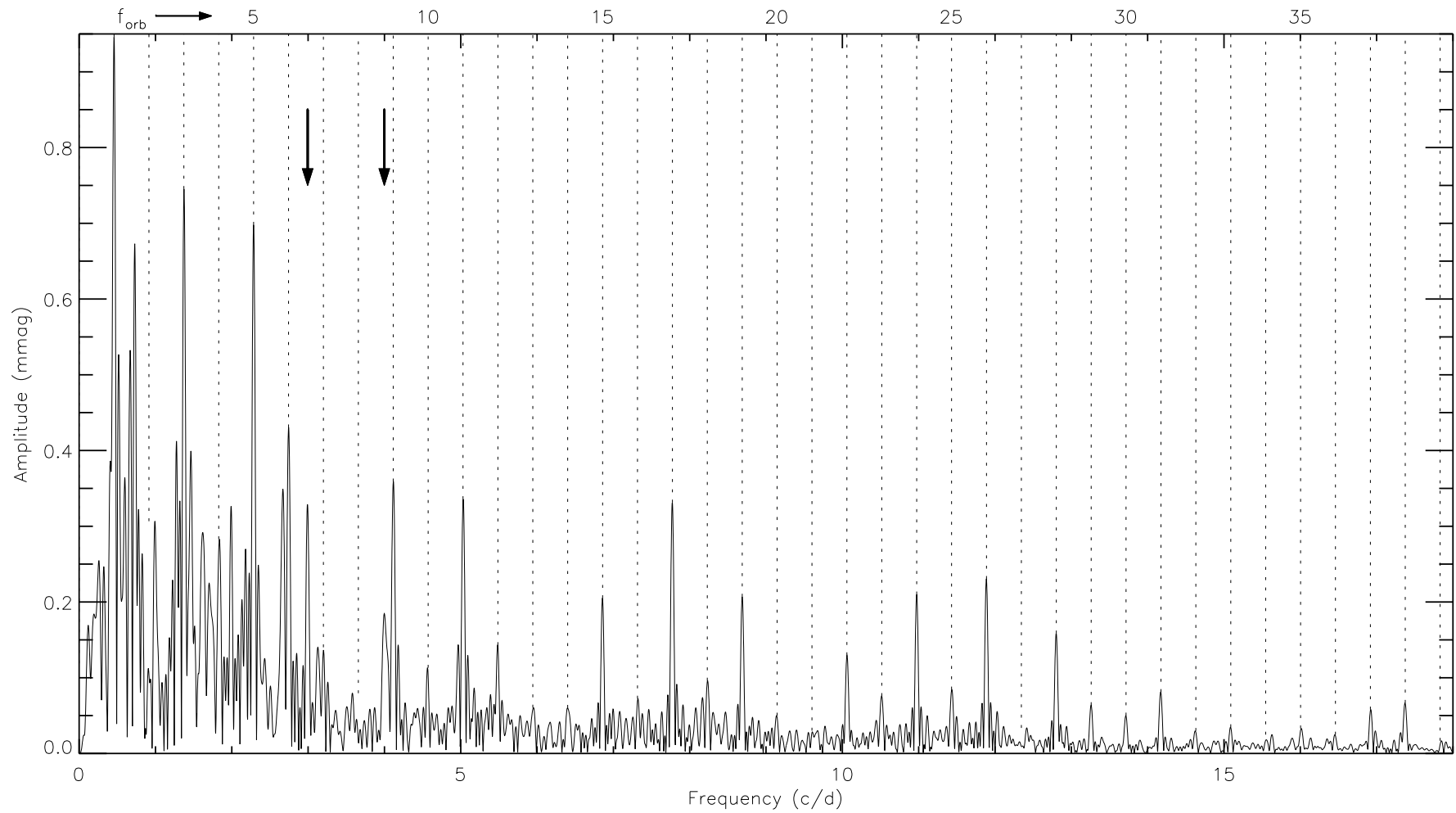}
\caption{\label{fig:freq} Amplitude spectrum of the \tess\ light curve of \targ\ after
subtraction of the {\sc jktebop} binary model. The dotted lines indicate multiples of
the orbital frequency, and the multiplicative factor is given above the top of the plot
for each fifth integer. The arrows indicate the two frequencies that are significantly
detected and do not coincide with orbital harmonics.} \end{sidewaysfigure}

Fig.~\ref{fig:freq} shows the amplitude spectrum of the residuals of the {\sc jktebop} fit to the unbinned SAP data. A large number of frequencies are present at multiples of the orbital frequency ($f_{\rm orb} = 0.45724$~d$^{-1}$). The highest-frequency of these is at 60$f_{\rm orb}$ (not shown on the plot). The two significant frequencies are indicated in Fig.~\ref{fig:freq} using arrows. The available data are not sufficient to indicate which of the components these frequencies arise from. This can sometimes be determined by looking at the residuals of the fit to the eclipses, but in the case of \targ\ the variability during eclipse is dominated by the signals at multiples of the orbital frequency.


\subsection*{A Roche-geometry model of the light curve}

The light curve binned into 400 points in orbital phase was modelled using the Wilson-Devinney code \cite{WilsonDevinney71apj,Wilson79apj}, in order to determine the photometric properties of the system. We used the 2004 version of the code ({\sc wd2004}) driven by the {\sc jktwd} wrapper \cite{Me+11mn}. {\sc wd2004} uses Roche geometry to accurately model the light curves of distorted stars in close binary systems.

\begin{figure}[t] \centering \includegraphics[width=\textwidth]{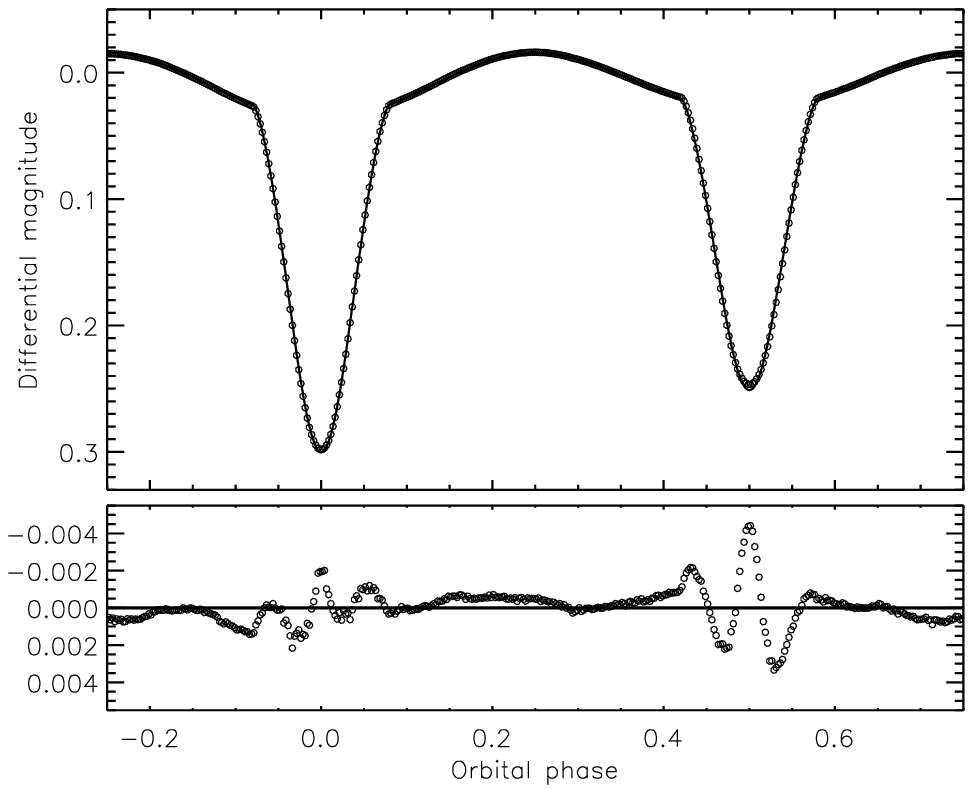} \\
\caption{\label{fig:phase} Best fit to the binned light curve of \targ\ using {\sc wd2004}.
The phase-binned data are shown using open circles and the best fit with a continuous line.
The residuals are shown on an enlarged scale in the lower panel.} \end{figure}

We have not been able to get a good fit to the light curve. Our best attempt is shown in Fig.\,\ref{fig:phase} and has significant residuals, as large as 4~mmag, through both eclipses. The reason for this remains unclear, as we have not encountered this precise problem in previous work on similar or different binary systems \cite{Me20obs,MeBowman22mn,MeClausen07aa,Pavlovski+09mn,Pavlovski++18mn}. In what follows we discuss the default setup of the modelling process and our attempts to improve the fit.

\textit{Operation mode}. The two relevant choices in {\sc wd2004} are mode 0 (\Teff\ values and light contributions are not forced to be consistent) and mode 2 (the \Teff\ values and passband-specific light contributions are forced to be consistent via the use of tabulated predictions from stellar model atmospheres). In our default approach we used mode 0, fixed the \Teff\ values at those from W09, and directly fitted for the contributions of the two stars to the total light of the system. As an alternative we chose mode 2 and fitted for the \Teff\ of star B, finding a very similar value (18\,424 versus 18\,500~K) and a similar quality of fit. We used a Johnson $R$ passband as the closest available option to the \tess\ band, but obtained practically identical results with a Johnson $I$ passband.

\textit{Numerical precision.} Our inital solutions were performed with a low numerical precision of \texttt{N1=N2=30} (see the {\sc wd2004} user guide \cite{WilsonVanhamme04}). Increasing this to the maximum value of 60 allowed a slight improvement in the fit.

\textit{Mass ratio}. We fixed the mass ratio at the value of $0.695 \pm 0.003$ measured spectroscopically by W09. The alternative approach of fitting for this quantity gives a better fit to the light curve but a mass ratio, 0.539, which is quite discrepant with the spectroscopic value.

\textit{Orbital eccentricity.} \targ\ is expected to have a circular orbit due to its age (W09) and short tidal timescales \cite{Zahn75aa,Zahn77aa}. However, a small eccentricity does give more freedom to fit eclipse profiles because it allows the primary and secondary eclipses to have different impact parameters. If the argument of periastron, $\omega$, is set to 90\degr\ or 270\degr\ it is possible to have an eccentric orbit where the secondary eclipse is still at phase 0.5. We tried this but were unable to significantly improve the fit. All solutions prefer at most a small eccentricity (0.01 or less) and there is no clear evidence for non-circularity.

\textit{Rotation rate.} Tidal effects are also expected to have caused the rotation of the stars to synchronise with the orbital motion, so in our default solution we assumed synchronous rotation. Attempts to fit the rotation rate of star~B failed because it has a negligible effect on the shape of the light curve. Fitting for the rotation rate of star~A yielded a determinate solution for faster rotation (1.7 times the synchronous value) but an only slightly improved fit. W09 measured the rotational velocities of the two stars from their spectral line profiles, finding them to be consistent with synchronous rotation.

\begin{figure}[t] \centering \includegraphics[width=\textwidth]{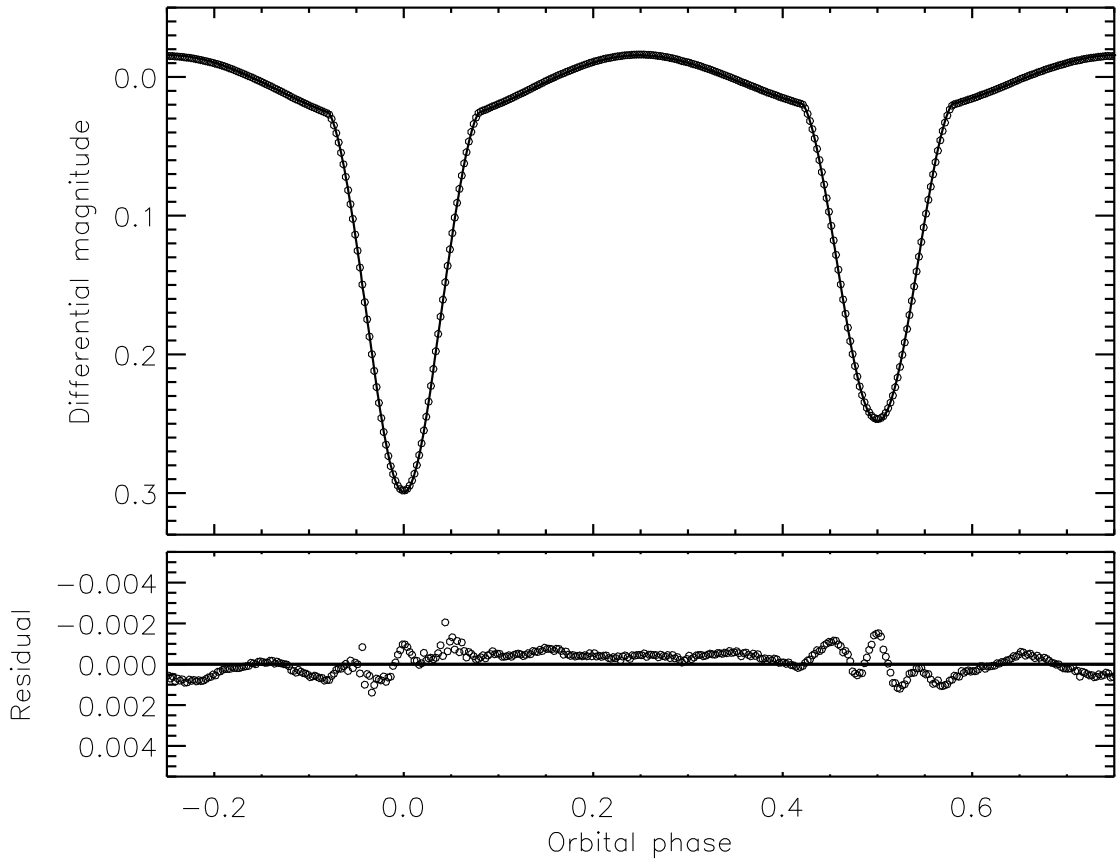} \\
\caption{\label{fig:albedo} Same as for Fig.~\ref{fig:phase} except for a solution where
the albedos of the two stars were fitted.} \end{figure}

\textit{Albedo.} We were able to find a significantly better model of the \tess\ light curve by fitting for the albedos of the two stars. The residuals decreased from 0.96~mmag to 0.51~mmag and the residuals during eclipse became much smaller (Fig.\,\ref{fig:albedo}). However, the fitted values of the albedo are 4.8 and 4.5, so are physically unrealistic \cite{Claret01mn}. Such a large albedo would require the stars to emit much more light in the \tess\ passband than is incident on them. We have seen a similar problem in the past (e.g.\ KIC 10661783 \cite{Me+11mn,Lehmann+13aa,Me+20mn}) and suspect that it arises due to a physical effect not included in the {\sc wd2004} model. For our default values we therefore fixed the albedos of both stars to 1.0.

\textit{Gravity darkening.} Both components are expected to have a gravity darkening exponent of $\beta = 1.0$ (Ref.~\cite{Claret98aas}). A significantly improved fit can be obtained by specifying values very different from this; a straight fit returns $\beta = -7.7$ for star~B, which must be rejected on physical grounds.

\textit{Reflection effect.} {\sc wd2004} includes the option of a detailed treatment of the reflection effect \cite{Wilson90apj} which we normally do not use because of the additional computing time needed. Use of the detailed treatment caused no improvement in the quality of the fit.

\textit{Third light.} A small amount of third light is expected to be common in data from \tess\ because of the large pixel size (21\as). We included this parameter by default and consistently obtained a small negative value. This is physically plausible if the background subtraction in the data reduction process is imperfect, and has been seen before in a similar system \cite{Vanreeth+22aa}.

\textit{Limb darkening} (LD). Our original fits used the linear LD law and coefficients fixed at values interpolated from the tables of Van Hamme \cite{Vanhamme93aj}. Use of the square-root or logarithmic LD laws gave almost identical results. Fitting for one LD coefficient per star allowed a small improvement in the fit. For our final fits we assumed logarithmic LD and fitted for the coefficients.

\textit{Inverted ratio of the radii.} It is often possible to get relatively good fits to the light curve of a dEB for both a certain ratio of the radii ($k = r_{\rm B}/r_{\rm A}$) and its reciprocal ($1/k$). We were indeed able to locate a solution with $k > 1$ for \targ, but the quality of the fit was significantly worse and the values of the fitted LD coefficients were very different from theoretical predictions based on stellar model atmospheres. The $k > 1$ solution is also discrepant with the spectroscopic light ratio measured by W09.

\textit{Data reduction.} We obtained a fit to the PDCSAP data to compare to the default solution using the SAP data. The measured parameter values and residuals were almost identical between the two fits.

\textit{Subsections of the light curve.} We fitted the sector 43 and sector 45 light curves separately, finding that the two solutions were extremely similar. The behaviour of \targ\ is therefore consistent over a time interval of at least 76~d.

\textit{Light curve model.} The {\sc jktebop} code was used to fit the phased light curve, to provide a comparison to the {\sc wd2004} solution. In our experience {\sc jktebop} can obtain good fits even beyond the limits of its applicability \cite{Me+11mn}. We were not able to get a better fit without allowing the code to use unphysical values for the limb darkening coefficients and reflection effect.


\subsection*{Final model of the light curve}

In light of the problems outlined above, we abandoned our attempts to get a \emph{good} fit to the light curve and settled for merely the \emph{best} fit, from the point of view of low residuals whilst remaining physically reasonable. Our final parameter values are based on fitting the phase-binned sector 43 light curves using {\sc wd2004} in mode 0, the simple reflection effect, the maximum numerical grid size of 60, a circular orbit, and the logarithmic LD law. The fitted parameters were the potentials of the two stars, the orbital inclination, the linear LD coefficient for each star, the phase of primary minimum, the light contributions of the two stars, and third light. The results are given in Table~\ref{tab:wd}.

\begin{table} \centering
\caption{\em Summary of the parameters for the {\sc wd2004} solution of the \tess\ light curve of \targ. Uncertainties
are only quoted when they have been assessed by comparison between a full set of alternative solutions. \label{tab:wd}}
\begin{tabular}{lcc}
{\em Parameter}                           & {\em Star A}          & {\em Star B}          \\[3pt]           
{\it Control parameters:} \\
{\sc wd2004} operation mode               & \multicolumn{2}{c}{0}                         \\                
Treatment of reflection                   & \multicolumn{2}{c}{1}                         \\                
Number of reflections                     & \multicolumn{2}{c}{1}                         \\                
Limb darkening law                        & \multicolumn{2}{c}{2 (logarithmic)}           \\                
Numerical grid size (normal)              & \multicolumn{2}{c}{60}                        \\                
Numerical grid size (coarse)              & \multicolumn{2}{c}{60}                        \\[3pt]           
{\it Fixed parameters:} \\
Mass ratio                                & \multicolumn{2}{c}{0.695}                     \\                
Orbital eccentricity                      & \multicolumn{2}{c}{0.0}                       \\                
Rotation rates                            & 1.0                   & 1.0                   \\                
Bolometric albedos                        & 1.0                   & 1.0                   \\                
Gravity darkening                         & 1.0                   & 1.0                   \\                
\Teff\ values (K)                         & 20\,500               & 18\,500               \\                
Bolometric linear LD coefficient          & 0.5494                & 0.5760                \\                
Bolometric logarithmic LD coefficient     & 0.2339                & 0.2184                \\                
Passband logarithmic LD coefficient       & 0.5124                & 0.4881                \\[3pt]           
{\it Fitted parameters:} \\
Phase shift                               & \multicolumn{2}{c}{$-0.00003 \pm 0.00002$}    \\                
Potential                                 & $3.843 \pm 0.030$     & $4.788 \pm 0.066$     \\                
Orbital inclination (\degr)               & \multicolumn{2}{c}{$77.50 \pm 0.33$}          \\                
Light contributions                       & $10.11 \pm 0.27$      & $2.90  \pm 0.11$      \\                
Passband linear LD coefficient            & $0.24 \pm 0.14$       & $0.12 \pm 0.20$       \\                
Third light                               & \multicolumn{2}{c}{$-0.032 \pm 0.023$}        \\                
{\it Derived parameters:} \\
Light ratio                               & \multicolumn{2}{c}{$0.287 \pm 0.013$}         \\                
Fractional radii                          & $0.3242 \pm 0.0022$   & $0.1923 \pm 0.0034$   \\[10pt]          
\end{tabular}
\end{table}

To determine the uncertainties in the fitted parameter values we ran new fits with different input parameters or approachs. These fits comprised: changing the mass ratio by $\pm$0.003; changing the rotation rate by $\pm$0.1, changing the albedo by $\pm$0.1; changing the gravity darkening exponents by $\pm$0.1; a numerical grid size of 40 instead of 60, use of mode 2 instead of 0, use of the detailed reflection effect; and linear instead of logarithmic LD. For each we calculated the changes in the values of the fitted parameters, then added these changes in quadrature to obtain the full errorbars for the parameters. We also considered fits with the LD coefficients fixed at the theoretical values instead of fitted for, and with third light fixed at zero, but in both cases the residuals were significantly higher so we did not include the results in the errorbars. The uncertainties reported by the differential-corrections fitting algorithm in {\sc wd2004} are in all cases much smaller than the uncertainties quoted in Table~\ref{tab:wd}.

The fractional radii, $r_{\rm A}$ and $r_{\rm B}$, are the most useful results in Table~\ref{tab:wd}. $r_{\rm A}$ is highly consistent between alternative fits so is precisely determined. $r_{\rm B}$ varies more, the biggest differences being seen for numerical precision and a change of gravity darkening exponent. As we were not able to get a good fit to the data, the uncertainty in $r_{\rm A}$ in Table~\ref{tab:wd} should be seen as a lower limit. In the following analysis we accounted for this by doubling the errorbar. It was not necessary to do the same for $r_{\rm B}$ because its errorbar was already significantly larger. In the analysis below we explicitly assume that the values of $r_{\rm A}$ and $r_{\rm B}$ are reliable even though we were not able to get a good fit to the light curve.

The light ratio in Table~\ref{tab:wd}, $\ell_2/\ell_1 = 0.287 \pm 0.013$, is in excellent agreement with the spectroscopic value of $0.25 \pm 0.05$ determined by W09. The photometric light ratio is also expected to be be slightly higher than the spectroscopic value because the TESS passband (approximately 590--990~nm) is redder than the blue wavelength range covered by W09's spectra (425--457~nm).


\section*{Physical properties of \targ}

We have calculated the physical properties of the \targ\ system using the $r_{\rm A}$, $r_{\rm B}$ and orbital inclination from Table~\ref{tab:wd}, with the errorbar on $r_{\rm A}$ doubled. To this we added the period found above and velocity amplitudes of $K_{\rm A} = 151.4 \pm 0.3$~km~s$^{-1}$ and $K_{\rm A} = 217.9 \pm 1.0$~km~s$^{-1}$ from W09. The calculations were performed with the {\sc jktabsdim} code \cite{Me++05aa}. To determine the distance to the system we used the apparent magnitudes given in Table~\ref{tab:info}, bolometric corrections from Girardi \etal\ \cite{Girardi+02aa}, and an interstellar extinction of $E(B-V) = 0.18 \pm 0.09$\,mag from the {\sc stilism}\footnote{\texttt{https://stilism.obspm.fr}} online tool \cite{Lallement+14aa,Lallement+18aa}. The results are given in Table~\ref{tab:absdim}.

\begin{table} \centering
\caption{\em Physical properties of \targ\ defined using the nominal solar units given by IAU 2015 Resolution
B3 (Ref.\ \cite{Prsa+16aj}). The \Teff\ values are 1500~K lower than those from W09. \label{tab:absdim}}
\begin{tabular}{lr@{\,$\pm$\,}lr@{\,$\pm$\,}l}
{\em Parameter}        & \multicolumn{2}{c}{\em Star A} & \multicolumn{2}{c}{\em Star B}    \\[3pt]
Mass ratio                                  & \multicolumn{4}{c}{$0.6948 \pm 0.0035$}       \\
Semimajor axis of relative orbit (\Rsunnom) & \multicolumn{4}{c}{$16.352 \pm 0.051$}        \\
Mass (\Msunnom)                             &  7.237  & 0.078       &  5.028  & 0.038       \\
Radius (\Rsunnom)                           &  5.301  & 0.074       &  3.144  & 0.056       \\
Surface gravity ($\log$[cgs])               &  3.849  & 0.012       &  4.144  & 0.015       \\
Density ($\!\!$\rhosun)                     &  0.0486 & 0.0020      &  0.1617 & 0.0086      \\
Synchronous rotational velocity ($\!\!$\kms)&  122.6  & 1.7         &  72.7   & 1.3         \\
Effective temperature (K)                   &   19000 & 1000        &   17000 & 1000        \\
Luminosity $\log(L/\Lsunnom)$               &   3.650 & 0.044       &   3.018 & 0.049       \\
$M_{\rm bol}$ (mag)                         &$-$4.39  & 0.11        &$-$2.81  & 0.12        \\
Distance (pc)                               & \multicolumn{4}{c}{$759 \pm 23$}              \\[3pt]
\end{tabular}
\end{table}

We have measured the masses and radii of the two stars to precisions of 1.8\% or better, so they are suitable for inclusion in the Detached Eclipsing Binary Catalogue (\debcat\footnote{\texttt{https://www.astro.keele.ac.uk/jkt/debcat/}}, Ref. \cite{Me15debcat}). We find slightly smaller masses than did W09 ($7.24 \pm 0.08$ and $5.03 \pm 0.04$\Msun\ versus $7.42 \pm 0.08$ and $5.16 \pm 0.03$\Msun), but only because our measurement of the orbital inclination is higher. More interestingly, our radius measurements are much lower: $5.30 \pm 0.07$ and $3.14 \pm 0.06$\Rsun\ versus W09's $5.60 \pm 0.04$ and $3.76 \pm 0.03$\Rsun. Our results are based on a \tess\ light curve incomparably better than the ASAS data available to W09, so should be preferred; the errorbars in the published measurements look too small. Although we were not able to find a good fit to the \tess\ data, the residuals are much smaller than the scatter of the ASAS data so it is reasonable to expect that both datasets suffer from the problem. Our preferred distance measurement is from the $K_s$ band, and is consistent with those from the other bands. Our initial distance measurement (see below) was $791 \pm 24$~pc, 1.0$\sigma$ longer than the parallax-based distance of $757 \pm 25$~pc from \gaia\ EDR3.

\begin{figure}[t] \centering \includegraphics[width=\textwidth]{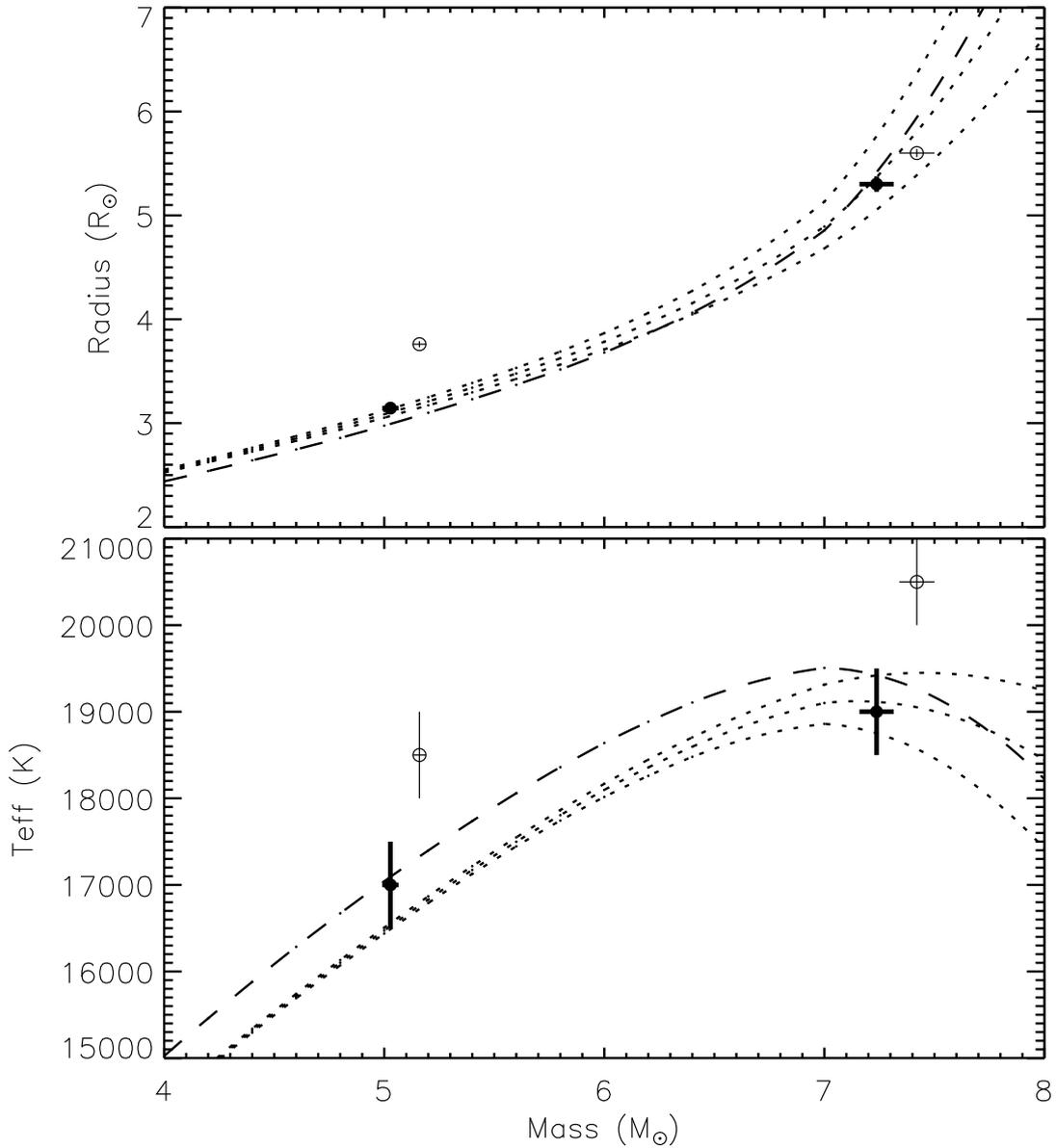} \\
\caption{\label{fig:theo} Mass--radius and mass--\Teff\ plots showing the properties
of \targ\ versus predictions from the {\sc parsec} models \cite{Bressan+12mn}.
The dotted lines are model predictions for $Z=0.020$ and ages of 31, 33 and 35 Myr.
The dashed lines are predictions for $Z=0.014$ and an age of 35\,Myr. The filled
circles with thick errorbars show the properties determined in the current work.
The open circles with thin errorbars show the values found by W09.} \end{figure}

To investigate the discrepancy between the published and our own measurements of the properties of the stars, we have compared them to theoretical predictions from the {\sc parsec} stellar evolutionary code \cite{Bressan+12mn}. We did this in mass--radius and mass--\Teff\ parameter space as these are good diagnostic plots \cite{MeClausen07aa}. We find that the masses and radii of the stars are matched by predictions for an age of $33 \pm 2$~Myr and a heavy-element abundance of $Z=0.020$. This is probably consistent with solar abundance considering recent developments \cite{GrevesseSauval98ssr,Asplund+09araa,Magg+22aa}. Their \Teff\ values are higher than predicted by approximately 1500~K, which corresponds to slightly less than the difference between the B2\,V and B2.5\,V spectral type \cite{PecautMamajek13apjs}. If we lower the \Teff\ values by this amount, we find a distance to the system of $759 \pm 23$~pc that is in almost perfect agreeement with \gaia\ EDR3. This represents our preferred set of system parameters as specified in Table~\ref{tab:absdim}. A comparison with other similar sets of theoretical models \cite{Pols+98mn,Pietrinferni+04apj} led to the same conclusions. The comparison is shown graphically in Fig.\ref{fig:theo}.

We then performed the same comparison but with the system properties from W09. We found a rough agreement for $Z=0.060$ and an age of 22~Myr, but even at this extremely high metallicity (the highest available in {\sc parsec}) the models underpredicted the radius and \Teff\ of star~B by approximately 3$\sigma$ and 2$\sigma$, respectively. We conclude that the published properties for this system are inconsistent with current stellar evolutionary models. W09 showed a Hertzsprung-Russell diagram (their fig.~7) in which the stellar properties agreed with the Lejeune \& Schaerer \cite{LejeuneSchaerer01aa} isochrones for an age of 25~Myr, but the stars were significantly too massive to match the predictions from Schaller \etal\ \cite{Schaller+92aas} (a situation also noticed by W09). This highlights how a decent agreement in a HR diagram is not sufficient for dEBs because mass is not directly involved; the use of mass--radius and mass--\Teff\ plots (Fig.~\ref{fig:theo}) is better as it allows conclusions to be drawn both more easily and more robustly.


\section*{Summary and conclusions}

\targ\ is a dEB containing two early-B stars with significant tidal distortions. We used the \tess\ light curve and published spectroscopic results to study the system. The residuals of the fit of an eclipsing binary model to the \tess\ data were used to obtain a frequency spectrum. We detected two significant pulsation frequencies in the system which could be due to $\beta$~Cephei or SPB star pulsation modes, making \targ\ one of the few known dEBs containing stars that show these types of pulsation. The frequency spectrum also shows a large number of peaks at integer multiples of the orbital frequency.

We modelled the \tess\ light curve using the {\sc wd2004} code to determine the photometric parameters of the system. We were unable to obtain a good fit and, after extensive investigation, do not know why. Possible answers are an effect of the pulsations, issues with the data reduction process, or the presence of surface inhomogeneities on the stars. Despite this, the best model of the \tess\ data is very well constrained and appears to be a significant improvement on previous work.

By making the assumption that the results of the Roche-model analysis above are reliable, and using published velocity amplitudes, we determined the physical properties of the component stars. These are in good agreement with theoretical predictions and with the \gaia\ EDR3 parallax for an approximately solar metal abundance and an age of 33~Myr, if a small decrease in the published \Teff\ values of the stars is applied. A previous analysis of the system (W09), using much more limited photometry, returned radius measurements in poor agreement with our own and with theoretical predictions. A new spectroscopic study of this system would be useful to check if the lower \Teff\ values we find are reasonable.


\section*{Acknowledgements}

We thank the referee for a positive and helpful report.
This paper includes data collected by the \tess\ mission and obtained from the MAST data archive at the Space Telescope Science Institute (STScI). Funding for the \tess\ mission is provided by the NASA's Science Mission Directorate. STScI is operated by the Association of Universities for Research in Astronomy, Inc., under NASA contract NAS 5–26555.
The following resources were used in the course of this work: the NASA Astrophysics Data System; the SIMBAD database operated at CDS, Strasbourg, France; and the ar$\chi$iv scientific paper preprint service operated by Cornell University.


\bibliographystyle{obsmaga}

\begin{thebibliography}{10}
\newcommand{\enquote}[1]{`(#1)'}

\bibitem{Andersen91aarv}
J.~{Andersen}, \textit{A\&ARv}, \textbf{3}, 91, 1991.

\bibitem{Torres++10aarv}
G.~{Torres}, J.~{Andersen} \& A.~{Gim{\'e}nez}, \textit{A\&ARv}, \textbf{18},
  67, 2010.

\bibitem{Me15aspc}
J.~{Southworth}, in \textit{Living Together: Planets, Host Stars and Binaries}
  (S.~M. {Rucinski}, G.~{Torres} \& M.~{Zejda}, eds.), 2015,
  \textit{Astronomical Society of the Pacific Conference Series}, vol. 496, p.
  321.

\bibitem{Demello++00apj}
D.~F. {de Mello}, C.~{Leitherer} \& T.~M. {Heckman}, \textit{ApJ},
  \textbf{530}, 251, 2000.

\bibitem{Robertson+10nat}
B.~E. {Robertson} \textit{et~al.}, \textit{Nature}, \textbf{468}, 49, 2010.

\bibitem{Langer12araa}
N.~{Langer}, \textit{ARA\&A}, \textbf{50}, 107, 2012.

\bibitem{Podsiadlowski+02apj}
P.~{Podsiadlowski}, S.~{Rappaport} \& E.~D. {Pfahl}, \textit{ApJ},
  \textbf{565}, 1107, 2002.

\bibitem{Podsiadlowski+04apj}
P.~{Podsiadlowski} \textit{et~al.}, \textit{ApJ}, \textbf{607}, L17, 2004.

\bibitem{Belczynski+20aa}
K.~{Belczynski} \textit{et~al.}, \textit{A\&A}, \textbf{636}, A104, 2020.

\bibitem{Chrimes+20mn}
A.~A. {Chrimes}, E.~R. {Stanway} \& J.~J. {Eldridge}, \textit{MNRAS},
  \textbf{491}, 3479, 2020.

\bibitem{Tkachenko+20aa}
A.~{Tkachenko} \textit{et~al.}, \textit{A\&A}, \textbf{637}, A60, 2020.

\bibitem{Johnston21aa}
C.~{Johnston}, \textit{A\&A}, \textbf{655}, A29, 2021.

\bibitem{Aerts++19araa}
C.~{Aerts}, S.~{Mathis} \& T.~M. {Rogers}, \textit{ARA\&A}, \textbf{57}, 35,
  2019.

\bibitem{Bowman+19aa}
D.~M. {Bowman} \textit{et~al.}, \textit{A\&A}, \textbf{621}, A135, 2019.

\bibitem{Bowman+20aa}
D.~M. {Bowman} \textit{et~al.}, \textit{A\&A}, \textbf{640}, A36, 2020.

\bibitem{Sana+14apjs}
H.~{Sana} \textit{et~al.}, \textit{ApJS}, \textbf{215}, 15, 2014.

\bibitem{Kobulnicky+14apjs}
H.~A. {Kobulnicky} \textit{et~al.}, \textit{ApJS}, \textbf{213}, 34, 2014.

\bibitem{Sana+12sci}
H.~{Sana} \textit{et~al.}, \textit{Science}, \textbf{337}, 444, 2012.

\bibitem{Me20obs}
J.~{Southworth}, \textit{The Observatory}, \textbf{140}, 247, 2020.

\bibitem{Walborn71apjs}
N.~R. {Walborn}, \textit{ApJS}, \textbf{23}, 257, 1971.

\bibitem{WalbornFitzpatrick90pasp}
N.~R. {Walborn} \& E.~L. {Fitzpatrick}, \textit{1990PASP..102..379W}.

\bibitem{Hipparcos97}
{ESA} (ed.), \textit{{The Hipparcos and Tycho catalogues. Astrometric and
  photometric star catalogues derived from the ESA Hipparcos space astrometry
  mission}}, \textit{ESA Special Publication}, vol. 1200, 1997.

\bibitem{Kazarovets+99ibvs}
E.~V. {Kazarovets} \textit{et~al.}, \textit{Information Bulletin on Variable
  Stars}, \textbf{4659}, 1, 1999.

\bibitem{Williams09aj}
S.~J. {Williams}, \textit{AJ}, \textbf{137}, 3222, 2009.

\bibitem{Pojmanski97aca}
G.~{Pojma{\'n}ski}, \textit{AcA}, \textbf{47}, 467, 1997.

\bibitem{Pojmanski02aca}
G.~{Pojma{\'n}ski}, \textit{AcA}, \textbf{52}, 397, 2002.

\bibitem{OroszHauschildt00aa}
J.~A. {Orosz} \& P.~H. {Hauschildt}, \textit{A\&A}, \textbf{364}, 265, 2000.

\bibitem{Hog+00aa}
E.~{H{\o}g} \textit{et~al.}, \textit{A\&A}, \textbf{355}, L27, 2000.

\bibitem{CannonPickering18anhar2}
A.~J. {Cannon} \& E.~C. {Pickering}, \textit{Annals of Harvard College
  Observatory}, \textbf{92}, 1, 1918.

\bibitem{Stassun+19aj}
K.~G. {Stassun} \textit{et~al.}, \textit{AJ}, \textbf{158}, 138, 2019.

\bibitem{Gaia21aa}
{Gaia Collaboration}, \textit{A\&A}, \textbf{649}, A1, 2021.

\bibitem{Cutri+03book}
R.~M. {Cutri} \textit{et~al.}, \textit{{2MASS All Sky Catalogue of Point
  Sources}} (The IRSA 2MASS All-Sky Point Source Catalogue, NASA/IPAC Infrared
  Science Archive, Caltech, US), 2003.

\bibitem{Ricker+15jatis}
G.~R. {Ricker} \textit{et~al.}, \textit{Journal of Astronomical Telescopes,
  Instruments, and Systems}, \textbf{1}, 014003, 2015.

\bibitem{Jenkins+16spie}
J.~M. {Jenkins} \textit{et~al.}, in \textit{Proc.\ SPIE}, 2016, \textit{Society
  of Photo-Optical Instrumentation Engineers (SPIE) Conference Series}, vol.
  9913, p. 99133E.

\bibitem{Me++04mn2}
J.~{Southworth}, P.~F.~L. {Maxted} \& B.~{Smalley}, \textit{MNRAS},
  \textbf{351}, 1277, 2004.

\bibitem{Me13aa}
J.~{Southworth}, \textit{A\&A}, \textbf{557}, A119, 2013.

\bibitem{Eastman++10pasp}
J.~{Eastman}, R.~{Siverd} \& B.~S. {Gaudi}, \textit{PASP}, \textbf{122}, 935,
  2010.

\bibitem{MeBowman22mn}
J.~{Southworth} \& D.~M. {Bowman}, \textit{MNRAS, submitted}, 2022.

\bibitem{Kurtz85mn}
D.~W. {Kurtz}, \textit{MNRAS}, \textbf{213}, 773, 1985.

\bibitem{StankovHandler05apjs}
A.~{Stankov} \& G.~{Handler}, \textit{ApJS}, \textbf{158}, 193, 2005.

\bibitem{Waelkens91aa}
C.~{Waelkens}, \textit{A\&A}, \textbf{246}, 453, 1991.

\bibitem{Aerts++10book}
C.~{Aerts}, J.~{Christensen-Dalsgaard} \& D.~W. {Kurtz},
  \textit{{Asteroseismology}} ({Astron.\ and Astroph.\ Library, Springer
  Netherlands, Amsterdam}), 2010.

\bibitem{WilsonDevinney71apj}
R.~E. {Wilson} \& E.~J. {Devinney}, \textit{ApJ}, \textbf{166}, 605, 1971.

\bibitem{Wilson79apj}
R.~E. {Wilson}, \textit{ApJ}, \textbf{234}, 1054, 1979.

\bibitem{Me+11mn}
J.~{Southworth} \textit{et~al.}, \textit{MNRAS}, \textbf{414}, 2413, 2011.

\bibitem{MeClausen07aa}
J.~{Southworth} \& J.~V. {Clausen}, \textit{A\&A}, \textbf{461}, 1077, 2007.

\bibitem{Pavlovski+09mn}
K.~{Pavlovski} \textit{et~al.}, \textit{MNRAS}, \textbf{400}, 791, 2009.

\bibitem{Pavlovski++18mn}
K.~{Pavlovski}, J.~{Southworth} \& E.~{Tamajo}, \textit{MNRAS}, \textbf{481},
  3129, 2018.

\bibitem{WilsonVanhamme04}
R.~E. {Wilson} \& W.~{Van Hamme}, \textit{Computing Binary Star Observables
  (Wilson-Devinney program user guide), available at
  ftp://ftp.astro.ufl.edu/pub/wilson}, 2004.

\bibitem{Zahn75aa}
J.~{Zahn}, \textit{A\&A}, \textbf{41}, 329, 1975.

\bibitem{Zahn77aa}
J.~{Zahn}, \textit{A\&A}, \textbf{57}, 383, 1977.

\bibitem{Claret01mn}
A.~{Claret}, \textit{MNRAS}, \textbf{327}, 989, 2001.

\bibitem{Lehmann+13aa}
H.~{Lehmann} \textit{et~al.}, \textit{A\&A}, \textbf{557}, A79, 2013.

\bibitem{Me+20mn}
J.~{Southworth} \textit{et~al.}, \textit{MNRAS}, \textbf{497}, L19, 2020.

\bibitem{Claret98aas}
A.~{Claret}, \textit{A\&AS}, \textbf{131}, 395, 1998.

\bibitem{Wilson90apj}
R.~E. {Wilson}, \textit{ApJ}, \textbf{356}, 613, 1990.

\bibitem{Vanreeth+22aa}
T.~{Van Reeth} \textit{et~al.}, \textit{A\&A}, \textbf{659}, A177, 2022.

\bibitem{Vanhamme93aj}
W.~{Van Hamme}, \textit{AJ}, \textbf{106}, 2096, 1993.

\bibitem{Me++05aa}
J.~{Southworth}, P.~F.~L. {Maxted} \& B.~{Smalley}, \textit{A\&A},
  \textbf{429}, 645, 2005.

\bibitem{Girardi+02aa}
L.~{Girardi} \textit{et~al.}, \textit{A\&A}, \textbf{391}, 195, 2002.

\bibitem{Lallement+14aa}
R.~{Lallement} \textit{et~al.}, \textit{A\&A}, \textbf{561}, A91, 2014.

\bibitem{Lallement+18aa}
R.~{Lallement} \textit{et~al.}, \textit{A\&A}, \textbf{616}, A132, 2018.

\bibitem{Prsa+16aj}
A.~{Pr{\v s}a} \textit{et~al.}, \textit{AJ}, \textbf{152}, 41, 2016.

\bibitem{Me15debcat}
J.~{Southworth}, in \textit{Living Together: Planets, Host Stars and Binaries}
  (S.~M. {Rucinski}, G.~{Torres} \& M.~{Zejda}, eds.), 2015,
  \textit{Astronomical Society of the Pacific Conference Series}, vol. 496, p.
  321.

\bibitem{Bressan+12mn}
A.~{Bressan} \textit{et~al.}, \textit{MNRAS}, \textbf{427}, 127, 2012.

\bibitem{GrevesseSauval98ssr}
N.~{Grevesse} \& A.~J. {Sauval}, \textit{Space Science Rev.}, \textbf{85}, 161,
  1998.

\bibitem{Asplund+09araa}
M.~{Asplund} \textit{et~al.}, \textit{ARA\&A}, \textbf{47}, 481, 2009.

\bibitem{Magg+22aa}
E.~{Magg} \textit{et~al.}, \textit{A\&A, in press, \texttt{arXiv:2203.02255}},
  2022.

\bibitem{PecautMamajek13apjs}
M.~J. {Pecaut} \& E.~E. {Mamajek}, \textit{ApJS}, \textbf{208}, 9, 2013.

\bibitem{Pols+98mn}
O.~R. {Pols} \textit{et~al.}, \textit{MNRAS}, \textbf{298}, 525, 1998.

\bibitem{Pietrinferni+04apj}
A.~{Pietrinferni} \textit{et~al.}, \textit{ApJ}, \textbf{612}, 168, 2004.

\bibitem{LejeuneSchaerer01aa}
T.~{Lejeune} \& D.~{Schaerer}, \textit{A\&A}, \textbf{366}, 538, 2001.

\bibitem{Schaller+92aas}
G.~{Schaller} \textit{et~al.}, \textit{A\&AS}, \textbf{96}, 269, 1992.

\end{thebibliography}

\end{document}